\documentclass[10pt,amsfonts]{article}

\usepackage{amssymb}
\usepackage{bm}
\usepackage{graphics}

\newcommand{\be}{\begin{equation}}
\newcommand{\ee}{\end{equation}}
\newcommand{\ba}{\begin{array}}
\newcommand{\ea}{\end{array}}
\newcommand{\bqa}{\begin{eqnarray}}
\newcommand{\eqa}{\end{eqnarray}}
\newcommand{\bea}{\begin{eqnarray}}
\newcommand{\eea}{\end{eqnarray}}

\begin{document}

\newtheorem{defi}{Definition}[section]
\newtheorem{lem}[defi]{Lemma}
\newtheorem{prop}[defi]{Proposition}
\newtheorem{theo}[defi]{Theorem}
\newtheorem{rem}[defi]{Remark}
\newtheorem{cor}[defi]{Corollary}

\newcommand{\qed}{\hfill $\Box$\vspace{.5cm}\medskip}

\title{Chern-Simons Gauge Invariance and Boundary Conformal Fields}

\author {Kumar Abhinav\footnote{E-mail: {\tt kumar.abhinav@bose.res.in}} and Samir K. Paul\footnote{E-mail: {\tt smr@bose.res.in}}\\$^1$S.N. Bose National Centre for Basic Sciences,\\ JD Block, Sector III, Salt Lake, Kolkata - 700106,  India \\}

\date{\today}
\maketitle

\hspace{1.20 in}

\vspace{.15 in}

\abstract{The topological non-Abelian Chern-Simons theory with a boundary is shown to require a scalar field companion in
order to preserve overall gauge-invariance both in the 3 dimensional manifold, as well as on its boundary. This scalar field, 
originally being a pure-gauge Chern-Simons excitation that indicates the most general topological gauge structure
in the bulk, is found to be crucial for {\it dynamic} scalar excitation on the 2 dimensional generic boundary manifold, 
the latter naturally representing a conformal mode. This gauge/CFT correspondence is general, for a compact gauge group.}

\bigskip


\bigskip

{\bf Keywords and Keyphrases :} Non-Abelian gauge theory, Chern-Simons theory, Conformal fields.

\section{Introduction}
Topological gauge theories have generated wide-spread interest for quite some time. Especially in 3 dimensions, such a
theory appear most naturally through Abelian $\left[A\wedge d A\right]$ or non-Abelian
$\left[T_r\left(A\wedge d A+\frac{2}{3}A\wedge A\wedge A\right)\right]$ Chern-Simons (CS) terms \cite{CST}, 
$A$ being a one-form. Both the versions implement a {\it gauge-invariant mass} \cite{TopM1,TopM11,TopM2,TopM22,Red,Red1,Boy,Boy1}
devoid of any symmetry-breaking mechanism, that further determines the spin of the gauge field \cite{TopM2,TopM22,Boy,Boy1}.
The geometrical origin of the non-Abelian CS term has been well-known to be the self-dual (anti-self dual) Yang-Mills 
action in 4 dimensions, {\it viz.}, 
$$T_r\,\int F^2:=T_r\,\int F\wedge F\equiv Tr\,\int d\left(A\wedge dA+\frac{2}{3}A\wedge A\wedge A\right),$$
with $F=dA+[A,A]=\pm~^*F$ being the field-strength two-form for $A$, so that the above bracket contains the {\it CS descendant}
in 3 dimensional space-time \cite{TJZW}. This CS term explicitly breaks parity and unlike the usual Maxwell term
$\left(F\wedge~^*F\right)$, does not effect any local dynamics, as is expected from the default topology. 
The physical origin of such a term is attributed to interactions with both fermions
\cite{TopM1,TopM11,TopM2,TopM22,Red,Red1,Boy,Boy1,Appel,H} as well as bosons \cite{PR01,pkp02} and these contributions
are protected beyond 1-loop \cite{CH}.

\paragraph*{}In addition to being the effective theory for a host of physical phenomena\footnote{Given the large number
of important publications in these areas, we are referring to comprehensive reviews/books.} such as quantum hall effect
\cite{Stone}, topological insulators \cite{Hasan} and fractional statistics \cite{Wilczek,Iengo} etc., the CS theory has been
linked deeply with planar gravity \cite{Pol01,W1,W2} and string theory \cite{W3,Marcos}. More specifically, the CS theory
is connected to two dimensional `rational' conformal field theories (CFTs), having a finite number of conformal blocks, 
which are in one-to-one correspondence with the CS Hilbert space \cite{W2,Seiberg,Lab,Poly,Nair}. All these approaches
consider pure CS theory on a 3 dimensional space-time of the form $\Sigma\times{\Bbb R}$, where $\Sigma$ is a compact 
Riemann surface. In absence of a {\it boundary}, the CS theory Hilbert space has one-to-one correspondence with the 
conformal blocks of a CFT defined on $\Sigma$. However, if $\Sigma$ has a boundary $\partial\Sigma$, the corresponding 
Hilbert space is infinite dimensional, with a chiral current algebra representation of the CFT on
$\partial\Sigma\times{\Bbb R}$ \cite{W2,Seiberg,Nair}. In the present work, we seek to obtain a boundary conformal 
dynamics, starting from a more relaxed condition, namely that of a CS theory defined in a generic 3 dimensional Riemannian manifold 
$M$ having a general 2 dimensional boundary $\partial M$. If the space is of the form $M=\Sigma\times{\Bbb R}$, with
boundary $\partial\Sigma$ of $\Sigma$, then $\partial M$ in our case will be $\partial\Sigma\times{\Bbb R}$.
The dynamical fields on the boundary are found to have representation in the bulk, as a part of the {\it generalized} Euclidean CS
action, completely justified by the corresponding topological (CS) hierarchy \cite{CST,TJZW}. Overall gauge-invariance plays 
the pivotal role in this process, with the scalar mode responsible for conformal dynamics on the boundary, additionally
identified as a pure-gauge solution corresponding to the {\it same} non-Abelian gauge group. The boundary contains a
non-dynamic, massive gauge excitation equivalent to chiral anomaly, for intrinsically flat manifolds like $\partial M=T^2$. However, for suitable gauge
choice on the boundary, a variety of $\partial M$s (including the erstwhile $\partial M=\partial\Sigma\times{\Bbb R}$ 
case) lead to conformal dynamics.

\paragraph*{}In the following, we consider the non-Abelian Euclidean CS action extended with a pure-gauge term in Section 2. Section 
3 deals with over-all gauge invariance of the system, including the boundary. Simple conformal structures are obtained 
on the boundary in Section 4, with various forms for manifolds with different topologies. Finally, we conclude.

\section{Non-Abelian Chern-Simons Theory}
We consider a general Riemannian 3-manifold $M$, not necessarily of the $\Sigma\times{\Bbb R}$ form, having a
2 dimensional boundary $\partial M$. Therein, a pure non-Abelian CS theory is defined through the action, 

\be
{\cal I}_{CS}^M=\kappa\int_M T_r\Big[A\wedge dA+i\frac{2}{3}gA\wedge A\wedge A\Big],\label{E1}
\ee
which exists essentially on the topology class of that manifold, thereby not affected by local geometry, so to say the
integrand is spread over the manifold. Herein, the one- and differential two-forms are defined respectively as,

\be
A:=A_\mu(x)dx^\mu \quad{\rm and}\quad dA=\partial_\mu A_\nu dx^\mu\wedge dx^\nu,\label{E2}, \quad \mu,\nu=1,2,3.
\ee 
The very form
of the action in Eq. \ref{E1} makes it topological, as for defining it on $M$ with a genuine curvature, there is no dependence
of the non-trivial metric $g_{\mu\nu}$ (geometry) is present. The trace in Eq. \ref{E1} is over the generators of the
gauge group, associated with the corresponding fields $A_\mu:=A^A_\mu T^A$.

\paragraph*{}From a topological point-of-view, ${\cal I}_{CS}^M$ belongs to a hierarchy of topological gauge structures 
across manifolds with different dimensionality. The present action is the {\it CS-descendant} of a
topological gauge Lagrangian, living in a 4 dimensional manifold $\tilde{M}$ \cite{TJZW},

\bea
{\cal I}_{D}^{\tilde{M}}=\frac{\kappa}{2}\int_{\tilde{M}}T_r\,\tilde{F}\wedge\tilde{F}&\equiv& \kappa T_r\,\int_{\tilde{M}}d\Big[A\wedge d A+i\frac{2}{3}gA\wedge A\wedge A\Big]\nonumber\\
&=&\kappa T_r\,\int_{\rm Boundary}\Big[A\wedge d A+i\frac{2}{3}gA\wedge A\wedge A\Big],\label{E8}
\eea 
where $\tilde{F}$ is the gauge field strength in 4 dimensions. The last equality above follows from Stokes
theorem. It is to be noticed that the second term 
above is topological in nature, as a two-form is wedged with itself, unlike the corresponding Maxwell term
$(\tilde{F}\wedge~^*\tilde{F})$. Being a total derivative, this leads to the CS Lagrangian on the 3 dimensional
{\it boundary} of the 4 dimensional space by Stokes theorem. However the most general CS descendant could be taken as,

\be
{\cal I}_{CS}^M=\kappa T_r\,\int_M \Big[A\wedge d A+i\frac{2}{3}gA\wedge A\wedge A-i\frac{2}{3}gd\Theta\wedge d\Theta\wedge d\Theta\Big],\label{E9}
\ee
where $\Theta(x)=\Theta^A(x)T^A$ is a non-Abelian scalar field such that $d\Theta$ transform as a gauge field.
Such a term can only exist in 3 dimensions, as in $\tilde{M}$ a term like,
$$T_r\,\int_{\tilde{M}}d\Big[d\Theta\wedge d\Theta\wedge d\Theta\Big],$$
vanishes identically. This kind of gauge field is found to be the source of conformal fields on the boundary of $M$. We
start with the {\it general} CS action in Eq. \ref{E9} on a 3-manifold $M$, which is {\it not} a boundary to any 4-manifold, but is
bounded itself by a 2 dimensional manifold $\partial M$. In principle one can add any number of scalar fields to the
above Lagrangian. However that does not lead to anything new except increased analytical complexity, especially regarding
the present work.

\paragraph*{}Important aspects of such a generalization to the CS theory can be derived from the corresponding dynamics,
given by the equations of motion (EOMs) for the two {\it disjoint} sectors:

\bea
&&\epsilon^{\alpha\mu\nu}\left[Tr\left(T^AT^B\right)\partial_\mu A_\nu^B+igTr\left(T^AT^BT^C\right)A_\mu^BA_\nu^C\right]=0,\nonumber\\
&&\epsilon^{\alpha\mu\nu}Tr\left(T^AT^BT^C\right)\partial_\alpha\left(\partial_\mu\Theta^B\partial_\nu\Theta^C\right)=0.\label{E3}
\eea
The $A$-sector is covariantly non-dynamical, being first order in derivatives, as is well-known. It can at most lead to
Galilean dynamics only in the presence of interactions \cite{H02}. For a given gauge group, one can always adopt the
normalization $Tr\left(T^AT^B\right)\propto\delta^{AB}$ and only the totally anti-symmetric part of $Tr\,\left(T^AT^BT^C\right)$
survives the absolute anti-symmetry in the covariant Greek indexes. Therefore, the first of the above equation reduces 
to,

\be
\epsilon^{\alpha\mu\nu}F_{\mu\nu}=0,\label{N02}
\ee
in the adjoint representation with a pure-gauge solution $A=-(i/g)U^{-1}dU$, so that it satisfies an EOM exactly
same as that for $\Theta$ above. This justifies considering $d\Theta$ to be a fundamental variable which is a 
pure gauge configuration. Such a configuration can undergo further gauge transformations. Moreover, the gauge transformation
acts in the space of gauge parameters, {\it independent} of the coordinate manifold $M$ wherein the partial derivative
is defined and $\Theta(x)$ transforms like a local scalar. Thus, the scalar sector is identified as an off-shell pure gauge
CS theory, with $d\Theta$ transforming identically under a generic gauge group defined by $U$ as $A$:

\be
A\to U^{-1}AU-\frac{i}{g}U^{-1}dU,\quad d\Theta\to U^{-1}d\Theta U-\frac{i}{g}U^{-1}dU\label{E13}
\ee
This aspect is of crucial importance, given that the non-Abelian CS action is {\it not} gauge-invariant in presence of a
finite boundary, as gauge transformation of the former yields a total derivative leading to a boundary term.

\subsection{Possible physical origin}
By its structure, the extended CS Lagrangian can physically be attributed to the {\it massive} matter
part of a generalized planar QCD Lagrangian \cite{BDP,Nakamura},

\bea
&&{\cal L}_{\rm Mat}=\bar{\Psi}_\alpha\left[ \begin{array}{cc} i\delta_{\alpha\beta}\gamma^\mu\partial_\mu-m\delta_{\alpha\beta}-g\gamma^\mu T_{\alpha\beta}^C A^C_\mu & 0 \\ 0 & i\delta_{\alpha\beta}\gamma^\mu\partial_\mu+m\delta_{\alpha\beta}-g\gamma^\mu T_{\alpha\beta}^C\partial_\mu\theta^C \end{array} \right]\Psi_\beta;\nonumber\\
&&\quad{\rm where}\quad \Psi_\alpha=\left( \begin{array}{c} \psi^a_\alpha \\ \psi^b_\alpha \end{array} \right),\label{E10}
\eea
with mass $m$ of two component $(a,~b)$ fermion field $\Psi_\alpha$ and coupling strength $g$, defined on $M$\footnote{Here,
to be able to consistently define fermions, $M$ needs to be without {\it intrinsic} curvature, as fermions cannot be
considered in a curved space-time.}. Here, $({\alpha,~\beta})$ denote the color species. In 3-space, the sign of $m$ also
doubles as that of the fermionic spin \cite{TopM2,TopM22,Boy,Boy1,Dunne} fixed for fermion and anti-fermion, and therefore
$A$ and $d\Theta$ couple to a particle-antiparticle pair. This is a crucial fact, as will be seen in the next section. 

\paragraph*{}In the effective gauge sector, apart from the usual wave-function renormalization, the presence of fermionic 
mass leads exactly to the CS contribution of Eq. \ref{E9}. The CS coefficient has the explicit expression:

\be
\kappa=(g^2/8\pi)(m/\vert m\vert),\label{N01}
\ee
that doubles as the topological {\it gauge mass} whose sign determines the gauge spin \cite{TopM2,TopM22,Boy,Boy1}. However,
this identification is restricted for $M$ being geometrically flat and the dynamic gauge (Maxwell) sector in the
effective theory will change the dynamics from that in Eq.s \ref{E3}\footnote{Even for ${\cal L}_{\rm Mat}$ being the 
complete Lagrangian, though only in the $A$-sector, the wave-function renormalization in the effective theory yields a
Maxwell-like contribution. It will dominate for small coupling $g\ll1$, which is required to attain an effective theory,
except in the long-wavelength limit \cite{TopM2,TopM22,H}.}. In the following, we consider the {\it primary} theory to 
be that in Eq. \ref{E9}, and forgo the identification of Eq. \ref{N01}.

\section{Complete Gauge Invariance with a Boundary}  
Under generic gauge transformation designated by $U$, common to both the gauge sectors on $M$, the two gauge fields
transform as in Eq. \ref{E13}. This changes the parent action of Eq. \ref{E9} as,

\be
{\cal I}_{CS}^M\to{\cal I}_{CS}^M+i\frac{\kappa}{g}T_r\int_Md\Big[dUU^{-1}\wedge\left(A-d\Theta\right)\Big].\label{E15}
\ee
The second term is the total derivative mentioned earlier, that does not vanish in presence of a finite boundary $\partial M$,
and therefore, the system looses gauge-invariance. In absence of the scalar field, along with the above boundary
contribution (without the $d\Theta$ term), a pure non-Abelian CS theory like that in Eq. \ref{E1} yields one more term
of the form,

\be
-\frac{\kappa}{3g^2}T_r\left(dUU^{-1}\wedge dUU^{-1}\wedge dUU^{-1}\right),\label{E14}
\ee 
under the same gauge transformation. This leads to a non-zero topological contribution (winding number) for large gauge
transformations at infinity \cite{TJZW,Dunne}
that changes the action by an integral phase, keeping the generating functional $\sim\left(\exp\,i{\cal I}\right)$ of
the system unchanged. However, in the present system with a finite $\partial M$, this creates a serious problem. As the
{\it off-shell} gauge field $A$ does not go to pure gauge there, there is no mapping of the 3-space $M$
into $S^3$, thereby violating off-shell gauge-invariance {\it even} in the bulk itself!. However, in the present model
of Eq. \ref{E9}, the equivalent contribution from the $d\Theta$ sector exactly cancels-out, preserving gauge-invariance
in the bulk. This is the most crucial aspect of the present model, and validates the presence of a second gauge field.
In geometrically flat $M$, this convenient aspect is attributed to the negative sign in front of the last term in Eq.
\ref{E9}, originating from the opposite sign of fermion masses in Eq. \ref{E10}. Thus, the interaction of the gauge-fields
with a fermion-antifermion doublet is a necessity here. However, this does {\it not} require the second gauge field to
be a pure-gauge one. This additional condition is necessary in order to have complete gauge-invariance of the system
including the boundary also. 

\paragraph*{}The aforementioned gauge non-invariance of Eq. \ref{E15} leaves a boundary contribution,

\be
i\frac{\kappa}{g}T_r\int_{\partial M}duu^{-1}\wedge\left(a-d\theta\right),\label{E16}
\ee
following Stokes theorem, on $\partial M$. This is a general property of any CS action, even an Abelian one \cite{Own}.
The fields (including the gauge element) on $\partial M$ are designated differently than their bulk counterparts, as the
boundary manifold may be different, both in geometry and topology. Also, the reduction of dimensions changes the
coordinate dependence in general, which can non-trivially effect the field configurations. Specifically $d\theta$ is
considered as the corresponding boundary counterpart of $d\Theta$. Accordingly, the gauge group is re-labeled as $u$,
with the gauge-group generators ($T^A$s) remaining the same. Same is true for the constants $g$ and $\kappa$.

\paragraph*{}The only way the overall gauge-invariance can be regained is by postulating an action of the form,

\be
{\cal I}_{B}^{\partial M}=\kappa T_r\int_{\partial M}d\theta\wedge a,\label{E17}
\ee
on the boundary manifold $\partial M$. Under a `new' set of gauge transformations,

\be
\rho\to u^{-1}\rho u-\frac{i}{g}u^{-1}du;\quad\rho=(a,~d\theta),\label{E18}
\ee
on $\partial M$, the boundary action transforms as,

\be
{\cal I}_{B}^{\partial M}\to{\cal I}_{B}^{\partial M}-i\frac{\kappa}{g}T_r\int_{\partial M}duu^{-1}\wedge\Big(a-d\theta-\frac{i}{g}duu^{-1}\Big).\label{E19}
\ee
The last integrand is the 2 dimensional analogue of that in Eq. \ref{E14}, that identically 
vanishes in two dimensions, confirming the absence of inherent topology therein \cite{TJZW}. Then the rest of the terms
exactly cancel the boundary contribution from the bulk in Eq. \ref{E16}. Therefore, for a 3 dimensional system with a finite
boundary, the complete gauge-invariant CS action is,

\bea
&&{\cal I}^{M+\partial M}_{CS}=\int_M{\cal L}_{CS}^M+\int_{\partial M}{\cal L}_{B}^{\partial M}\nonumber\\
&&\qquad\qquad\qquad\quad=\kappa\int_MT_r\Big[A\wedge d A+i\frac{2}{3}gA\wedge A\wedge A-i\frac{2}{3}gd\Theta\wedge d\Theta\wedge d\Theta\Big]\nonumber\\
&&\qquad\qquad\qquad\qquad+\kappa\int_{\partial M}T_r\Big[d\theta\wedge a\Big],\label{E20}
\eea
following respective transformation laws of Eq.s \ref{E13} and \ref{E18}. The corresponding gauge invariant EOMs for the
bulk sector are already given in Eq. \ref{E3}, leading to non-dynamic pure-gauge conditions, satisfied by both $A$ 
and $d\Theta$. On the boundary, the corresponding equations are,

\be
d a=0 \quad{\rm and}\quad \epsilon^{\mu\nu}\partial_\nu\theta=0,\label{E21}
\ee
leading to trivial, no-dynamic solutions.

\paragraph*{}The above results show why a field of the form $d\Theta$, that transforms as a gauge field under the 
same gauge group as $A$, was required in the bulk at the first place. Even in its absence, the standard pure CS theory of
Eq. \ref{E1} would yield a boundary term $i(\kappa/g)T_r\left(duu^{-1}\wedge a\right)$ that cannot
be compensated by any boundary Lagrangian. The additional winding-number-like term of Eq. \ref{E14} would
always have remained. This has to do with the non-Abelian nature of the gauge fields, as for the Abelian case, no such field
in the bulk is required \cite{V1}. However, the non-Abelian generalization naturally admits a pure-gauge field
of the form $d\Theta$ as a part of the topological descendant, as seen from Eq. \ref{E9}. All these requirements
cannot simultaneously be fulfilled by, say, a second CS term. The present Lagrangian of Eq.
\ref{E20} is the only one that provides overall gauge-independence over the whole of $M\oplus\partial M$. 

\paragraph*{}From the EOMs Eq. \ref{E21}, $a$ and $d\theta$ are of chiral nature. Indeed, the EOM of Eq. \ref{N02} essentially represented
a chiral current, and one would expect this property to be reflected in the boundary through overall gauge invariance.
We will see similar structure to be retained once we extend the boundary theory in order to incorporate dynamics therein.

\section{Gauge-Invariant Dynamics on the Boundary}
We intend to extend the gauge-invariant theory of Eq. \ref{E20} to include boundary dynamics. This is motivated
by the earlier observed scalar KdV solitons on such a boundary \cite{P,D} which appeared on the boundary as dual to planar gravity
\cite{W}. However, from a pure CS theory point-of-view, such a boundary dynamics has not been obtained for the non-Abelian
case. The simplest dynamic extension to the boundary Lagrangian in Eq. \ref{E17} can be taken as \cite{Own},

\be
{\cal L}_{B}^{\partial M}\to{\cal L}_{B}^{\partial M}=T_r\Big[\left(d\theta-a\right)\wedge~^*\left(d\theta-a\right)+\kappa d\theta\wedge a\Big].\label{E22}
\ee
The first term, containing the wedge with a Hodge dual is {\it not} topological, unlike the terms considered till now.
However, this term is form-invariant over the 2-manifold owing to the fact that a 2-manifold is conformally flat and the 
second term is also form-invariant as it is topological by construction. Thus, ${\cal L}_B^{\partial M}$ in Eq. \ref{E22}
is `spread over' the manifold in a form-invariant manner and it is therefore a possible Lagrangian on $\partial M$.
It is a valid Lagrangian density in 2-space also because it lives on the corresponding topology. Further, it is
invariant under the gauge transformation of Eq. \ref{E18}, and clearly leads to dynamics. Though $a$
is still not dynamic, it contributes critically through the equations of motions, which have the final forms,

\be
d a=0 \quad{\rm and}\quad \partial^2\theta=\partial_\mu a^\mu.\label{E23}
\ee
From the first equation, $a$ now represents the chiral mode, whereas the second
represents covariant dynamics of the {\it massless} boundary scalar mode $\theta$,
which becomes free in the covariant Lorentz gauge for $a$. However, such a naiv\'e gauge-fixing is not straight-forward
as $\theta$ also transforms under the same gauge group and so as the fields in the bulk. This topic will be considered 
later. In the present case, such a choice is too trivial as from the first equation above, $a$ is curl-less too. This
already hints at the possibility of constructing a linear combination of $\theta$ and $a$, that can represent a free
mass-less mode on $\partial M$, which is a conformal structure. 

\paragraph*{}Essentially, topological terms in the Lagrangian do not contribute to the dynamics of the system by their own,
owing to their essential non-locality. This becomes apparent in the Hamiltonian formalism, which is implicitly covariant,
wherein the the topological terms do not appear. It is a straight-forward evaluation through the Legendre transformation 
that the Hamiltonian corresponding to pure CS-type Lagrangian in Eq. \ref{E1} identically vanishes \cite{Dunne}.
However, this requires a-priori on-shell gauge-fixing. Following Ref. \cite{Dunne}, on considering the {\it Minkowskian counterpart}
of the present system with $\partial M$ being a 1+1 dimensional manifold, the Weyl gauge $a_0=0$ can be adopted.
The Hamiltonian corresponding to the dynamical Lagrangian of Eq. \ref{E22} will be non-vanishing.
As mentioned above, the intricacy of gauge-fixing in the present case means an {\it overall} gauge-fixing that manifests
simultaneously in the $d\theta$ sector and also in the bulk fields. Importantly, the boundary Hamiltonian will represent
a dynamics independent of the bulk, as $\partial M$ is {\it not} connected $M$ dynamically. In the Weyl gauge,
$a_0$ essentially is a Lagrange multiplier for the Gauss law constraint of the system, vanishing upon the Legendre
transformation. The Hamiltonian then obtained as,

\be
{\cal H}_{B}^{\partial M}=\frac{1}{2}T_r\left[\dot{\theta}^2-\left(\partial_1\theta-a_1\right)^2\right],\quad \dot{\theta}:=\partial_0\theta,\label{NE1}
\ee 
which does not contain any topological term. This result is quite general as it is independent of specifics of both $M$ and
$\partial M$. Although no explicit off-shell conformal structure appear until now, the presence of a dynamic
mass-less scalar in the Hamiltonian tempts in that direction, obstructed only by the interacting gauge field.

\paragraph*{}To obtain a conformal structure, specific reduction of the variables, based on the particular topological
structure of the system is in order. It is known \cite{W2,Marcos,Seiberg,Lab,Nair} that the boundary gauge field can be
resolved in longitudinal (L) and transverse (T) components as $a=a_L+a_T$. For $\partial M$, any vector can be cast in
the form \cite{Das},

\be
a=d\eta+^*d\xi, \quad \eta,~\xi\in{\Bbb R}\label{E24}
\ee
which owes to the fact that there are only two directions. This `splitting' of $a$ is generic to the particular
$\partial M$, so to say there is no parameter taking care of any other $\partial M$ diffeomorphic to the present one.
Any real shift in the `longitudinal' scalar $\eta$ corresponds to 
gauge transformation of $a$ \cite{Dunne}. For $\partial M$ being $T^2$, the decomposition has the generic form,

\be
{\cal A}=d_{\bar{z}}\chi+\Gamma(z;\tau)\alpha;\qquad {\cal A}=a_0+ia_1,\quad z=x_0+ix_1,\quad \chi=\eta-i\xi.\label{NE2}
\ee
The local function $\Gamma(z;\tau)$ depends on the topological parameter $\tau$, labeling $\partial M$. A {\it small} gauge 
transformation ({\it e. g.} $SU(N)$-type) effects exclusively the longitudinal part:
$\eta\to\eta+\lambda,~\lambda\in{\Bbb R}$, whereas a {\it large} gauge transformation shifts the constant parameter
$\alpha$\footnote{In a 3-manifold resolvable as $T^2\times{\Bbb R}$, $\alpha=\alpha(t)$. In our case, with
$\partial M=T^2$, it is a constant.} \cite{Dunne}. However, an orthogonal set of coordinates $(x_0,x_1)$ can always be
constructed on the 2-manifold $\partial M$ whether it is $R^2$, $T^2$ or $S^2$. Therefore the resolution in Eq.
\ref{E24} is quite valid.

\paragraph*{}As already had been mentioned, although a gauge field has no degrees of freedom in 2-space, the {\it same}
gauge constraint is spread over the whole field space defined in the bulk too. On attributing the gauge fixing exclusively
to $\partial M$, choosing a particular $u$ (or $U$) amounts for reduction of two degrees of freedom, leaving-out a single
physical one, as $\theta$ is a scalar and $a$ is a 2-vector. We will find this to be the case exactly. On substituting
Eq. \ref{E24} in Eq. \ref{E22}, the boundary Lagrangian takes the form,

\bea
&&{\cal L}_{B}^{\partial M}=T_r\Big[d\theta\wedge~^*d\theta+d\eta\wedge~^*d\eta-d\xi\wedge~^*d\xi
+2d\theta\wedge~^*d\left(\eta+\kappa\xi\right)\nonumber\\
&&\qquad\qquad\qquad+d\theta\wedge d\left(\xi+\kappa\eta\right)-d\eta\wedge d\xi\Big].\label{E25}
\eea
The kinetic part in $\xi$ represents negative-norm states, exactly canceling-out the positive-norm $\eta$ states. Thus,
there is no dynamical degree of freedom for the erstwhile $a$ field. This re-confirms the above inference regarding 
the constrained nature of the system. All three fields satisfy massless Klein-Gordon equation,

\be
(d~^*d) \zeta=0,\quad \zeta=\theta,~\eta,~\xi.\label{E26}
\ee
This re-confirms the prior gauge-fixing again. 
Simply stated, implementation of Eq. \ref{E24} reduces topological terms in Eq. \ref{E25} to total derivatives,
with no contribution as there is no boundary of a boundary. In both representations, we have only one dynamical field
($\theta$), with interactions, as the dynamical contributions of the other two cancel-out in the latter case. In the next,
it will be shown to be resolved into a free field theory.

\subsection{The Conformal Theory}
Even after non-trivial gauge-fixing, the boundary theory of Eq. \ref{E22} is left with interactions and topological terms,
preventing a conformal structure. The topological terms in Eq. \ref{E25} can be removed at the level of Lagrangian itself 
(off-shell) by performing a rotation in the space of field-variables defined on $\partial M$. This is a general property of
topological models like CS theory in the 3 dimension and BF theory in 4 dimensions\footnote{This point is not explicitly 
demonstrated in most literature. However, it can indirectly be seen in effective treatment of the said theories, leading to 
apparent non-local mass terms. In the present case though, no such non-locality arise, owing to the absence of the usual 
Maxwell dynamics at the tree level. Indeed, the gauge field is non-dynamical in the present case.}. For analytical brevity, 
we opt for the specific rotation accompanied with a gauge choice:

\be
d\tilde{\theta}:=d\theta+\kappa~^*a \quad{\rm and}\quad a:=d\tilde{\eta},\label{E28}
\ee
at the level of Eq. \ref{E22}, that corresponds to a unit functional Jacobian. The second expression amounts to a pure 
gauge form, like that in case of $d\theta$ ($d\Theta$), as $a$ satisfies $d a=0$ on-shell (Eq.s \ref{E23}). For
$\partial M=T^2$ this can correspond to a large gauge choice (${\cal X},~a=0$) \cite{Dunne}, as discussed before,
requiring judicious choice of coordinates. Essentially the system now becomes on-shell with overall gauge invariance of
the action being retained as before. As a result, the re-arranged Lagrangian is,

\be
{\cal L}_{B}^{\partial M}\equiv T_r\Big[d\left(\tilde{\theta}-\tilde{\eta}\right)\wedge~^*d\left(\tilde{\theta}-\tilde{\eta}\right)+\kappa^2d\tilde{\eta}\wedge~^*d\tilde{\eta}\Big]=T_r\Big[d\vartheta\wedge~^*d\vartheta+\kappa^2d\tilde{\eta}\wedge~^*d\tilde{\eta}\Big].\label{E29}
\ee
The above Lagrangian contains two free massless scalar modes $\vartheta=\tilde{\theta}-\tilde{\eta}$ and $\tilde{\eta}$,
representing conformal modes in a 2 manifold \cite{CFT1,CFT2} like $\partial M$. However, their obvious linear dependence
leads to only {\it one} independent dynamical mode, which is evident from the individual EOMs: 

\be
\partial^2\tilde{\theta}=\partial^2\tilde{\eta} \quad{\rm and}\quad \partial^2\tilde{\theta}=(1+\kappa^2)\partial^2\tilde{\eta}\qquad \left(\partial^2=\partial_\mu\partial^\mu\right),\label{E30}
\ee
respectively. Both relate d'Alembertians of the two fields, leading to the trivial solutions that both d'Alembertians vanish.
However, this result cannot be obtained unless one equation of motion is substituted in the other, containing the same 
variables. This justifies essentially one degree of freedom in the system as before, a signature of its constrained nature.
This exactly in accordance with $d\theta$ having one degree of freedom whereas $a_\mu$ has none. 

\paragraph*{}The manifestation of conformal dynamics in 2 space is ensured by local scale (Weyl) invariance. The later is
ensured by the vanishing of the trace of respective energy-momentum tensor \cite{CCJ},

\be
T_{\mu\nu}=2\frac{\delta{\cal L}}{\delta g^{\mu\nu}}-g_{\mu\nu}{\cal L}.\label{E31}
\ee 
For the boundary Lagrangian of Eq.s \ref{E22} and \ref{E25}, the respective traces,

\be
T^\mu_{~\mu}=2\kappa T_rd\theta\wedge a \quad{\rm and}\quad T^\mu_{~\mu}=2T_r\left[d\theta\wedge d\left(\xi+\kappa\eta\right)-d\eta\wedge d\xi\right],\label{E32}
\ee
turn out to be non-zero. However, they are exclusively of the topological nature whereas the energy-momentum tensor itself
is the generator of local translations. Moreover, in the second expressions, the integral over $\partial M$ do vanish as
it is a total derivative. As physical observables are essentially the integrals of these `densities', this leads to a 
vanishing result. However for conformal symmetry to exist, which is of local nature, the trace density itself must vanish.
This ambiguity is expectedly resolved with the Lagrangian representation of Eq. \ref{E29}, exactly leading to
$T^\mu_{~\mu}=0$\footnote{The energy-momentum tensor for free scalar field is {\it not} traceless in general, except in
2-space, allowing the latter case to be only example of conformal field theory with free
scalar field.}. Strictly speaking, this result ensures local scale or Weyl invariance of the theory
\cite{CCJ}, which means conformal invariance in a 2-manifold \cite{CFT1,CFT2}. 

\paragraph*{}A conformal structure can be obtained, however, without subjecting the
resolution in Eq. \ref{E24} in general. For a generic boundary 2-manifold $\partial M$, the first of Eq.s \ref{E28} is
always valid, leading to the {\it rotated} Lagrangian,

\be
{\cal L}_{B}^{\partial M}=\Big[\left(d\tilde{\theta}-a\right)\wedge~^*\left(d\tilde{\theta}-a\right)+\kappa^2a\wedge~^*a\Big].\label{E33}
\ee
The corresponding energy-momentum tensor is again traceless, ensuring Weyl symmetry. The field equations essentially yield
a mass-less scalar $\tilde{\theta}$, even without any gauge-fixing, with the $a$ sector being non-dynamic. Most
importantly, although the theory is not conformal except in ${\Bbb R}^2$ or in $T^2$ following large gauge transformation,
it is Weyl-invariant on a {\it generic} 2 dimensional $\partial M$. 

\paragraph*{}Eq. \ref{E33} embodies an interesting observation. The second term here is essentially a gauge mass term, and 
on inherently flat manifolds like ${\Bbb R}^2$ and $T^2$, appears as the celebrated chiral anomaly contribution
\cite{Das,Fujikawa}. This identification implies $\kappa^2=e^2/\pi$, $e$ being the coupling of $a$ with some
integrated-out massless fermions on $\partial M$. But since the gauge group on $M\oplus\partial M$ is uniquely parameterized
by $g$, $e=g$. Therefore, the interpretation of {\it effective chiral anomaly} is valid only if $\kappa^2=g^2/\pi$. However,
this is not possible if the CS term in $M$ is induced from a system like that in Eq. \ref{E10}, as it would 
impose $\kappa\sim g^2$ \cite{BDP,TopM2,TopM22,Boy,Boy1}, yielding a contradiction. 

\paragraph*{}On the other hand, this still leaves room for extension to ${\cal L}_{B}^{\partial M}$ by a massless boundary
fermion ($\sigma$) sector $\bar{\sigma}\left(i\gamma\cdot d-g\gamma\cdot a\right)\sigma$. When integrated-out, this sector 
manifests in a parametric shift: $\kappa^2\to\kappa^2+g^2/\pi$ in Eq. \ref{E33} due to chiral anomaly \cite{Das,Fujikawa}. 
It is to be noted that these results are exclusive to intrinsically flat manifolds, as fermions cannot be defined in
presence of curvature.

\paragraph*{}The present theory can be made {\it manifestly} conformal ({\it i. e.}, free scalar one) at the effective theory level
by integrating-out the non-dynamic gauge field. Up to a constant normalization, this results in complete Bosonization as,
\be
{\cal L}_{B}^{\partial M}=\left[\frac{5+\kappa^2}{1+\kappa^2}\right]T_rd\tilde{\theta}\wedge~^*d\tilde{\theta},\label{E34}
\ee 
which is conformal by construction on the 2-manifold $\partial M$. As Eq. \ref{E33} is valid in a generic $\partial M$, so
as the above conformal structure. This result is off-shell and effects in wave-function renormalization of the scalar
field due to the original interaction. However, the conformality of the full boundary theory is apparent through the
corresponding traceless energy-momentum tensor at an off-shell level.

\section{Conclusions}
To summarize the results, we have generalized the pure non-Abelian CS descendant to a generalized CS descendant.
Necessarily being of pure-gauge nature, this extension is decoupled from the usual gauge sector in $M$ and thus, does
not effect the characteristic CS dynamics of references \cite{W1,W2,W3,Marcos,Seiberg,Lab}, including the latter's
correspondence with 3 dimensional gravity. However, in presence of a finite boundary, overall gauge-invariance
necessarily couple the two fields in a topological manner on the boundary $\partial M$. In general, this allows for a
dynamic extension on $\partial M$, leaving only one degree of freedom, that encompasses $M\oplus\partial M$ and the
whole of the field space. 
\paragraph*{}Upon a rotation in field-space on $\partial M$, the free scalar mode separates from the non-dynamic gauge 
sector, irrespective of the geometric and topological nature of the prior. This makes the theory explicitly Weyl-invariant
and the dynamic scalar mode is identified as a conformal one on the boundary 2 manifold, and necessarily demands the boundary
gauge field $a$ also to be of pure-gauge form. For a generic $a$, conformal dynamics is obtained on {\it intrinsically flat}
boundary manifolds with additional interacting massless fermions, with the non-dynamic sector effectively attributed to
chiral anomaly. Indeed, the effective theory obtained by integrating the non-dynamical fields out, is a free scalar one.

\vspace{.15 in}

\noindent{\bf Acknowledgement:} KA appreciates fruitful discussions with Prof. P. K. Panigrahi and Dr. V. M. Vyas that
led to the basic idea of this work.


\begin{thebibliography}{99}
\bibitem{CST}S.-S. Chern and J. Simons, \emph{Characteristic forms and geometric invariants}, Ann. Math. {\bf 99} 18 (1974).

\bibitem{TopM1} J. F. Schonfeld, \emph{A mass term for three-dimensional gauge fields}, {\it Nucl. Phys. B} {\bf 185} (1981) 157.
\bibitem{TopM11} I. Affleck, J. Harvey and E. Witten, \emph{Instantons and (super-) symmetry breaking in (2+1) dimensions}, {\it Nucl. Phys. B} {\bf 206} (1982) 413.
\bibitem{TopM2} S. Deser, R. Jackiw and S. templeton, \emph{Three-Dimensional Massive Gauge Theories}, {\it Phys. Rev. Lett.} {\bf 48} (1982) 975. 
\bibitem{TopM22}S. Deser, R. Jackiw and S. templeton, \emph{Topologically massive gauge theories}, {\it Ann. Phys.} {\bf 140} (1982) 372.
\bibitem{Red} A. Redlich, \emph{Gauge Noninvariance and Parity Nonconservation of Three-Dimensional Fermions}, {\it Phys. Rev. Lett.} {\bf 52}, 18 (1984).
\bibitem{Red1}A. Redlich, \emph{Parity violation and gauge noninvariance of the effective gauge field action in three dimensions}, {\it Phys. Rev. D} {\bf 29}, 2366 (1984). 
\bibitem{Boy} D. Boyanovsky, R. Blankenbecler and R. Yahalom, \emph{Physical origin of topological mass in 2 + 1 dimensions}, {\it Nucl. phys. B} {\bf 270} (1986) 483.
\bibitem{Boy1}B. Binegar, \emph{Relativistic field theories in three dimensions}, {\it J. Math. Phys.} {\bf 23} (1982) 1511.
\bibitem{TJZW}{\em Current Algebra and Anomalies}, S. B.Treiman, R. Jackiw, B. Zumino and E. Witten, Princeton University Press, Princeton, 1985.
\bibitem{Appel}T. W. Appelquist, M. Bowick, D. Karabali, and L. C. R. Wijewardhana, \emph{Spontaneous chiral-symmetry breaking in three-dimensional QED}, {\it Phys. Rev. D} {\bf 33} (1986) 3704.
\bibitem{H} C. R. Hagen, \emph{A new gauge theory without an elementary photon}, {\it Ann. Phys.} {\bf 157} (1984) 342.
\bibitem{PR01}R. Pisarski and S. Rao, {\it Topologically Massive Chromodynamics in the Perturbative Regime}, Phys. Rev. D {\bf 32}, 2081 (1985).
\bibitem{pkp02}C. R. Hagen, P. Panigrahi, and S. Ramaswamy, \emph{Still More Corrections to the Topological Mass}, {\it Phys. Rev. Lett.} {\bf 61} (1988) 389.
\bibitem{CH}S. Coleman and B. Hill, \emph{No more corrections to the topological mass term in $QED_3$}, {\it Phys. Lett. B} {\bf 159} (1985) 184.
%\bibitem{Ren} T. Appelquist and R. Pisarski, \emph{High-temperature Yang-Mills theories and three-dimensional quantum chromodynamics}, {\it Phys. Rev. D} {\bf 23} (1981) 2305.
%\bibitem{Inf}T. Appelquist and U. Heinz, \emph{Three-dimensional $O(N)$ theories at large distances}, {\it Phys. Rev. D} {\bf 24} (1981) 2169.
\bibitem{Stone}M. Stone, {\em Quantum Hall Effect} (World Scientific, Singapore, 1992).
%\bibitem{Neto}A. H. C. Neto, F. Guinea, N. M. R. Peres, K. S. Novoselov and A. K. Geim, \emph{The electronic properties of graphene},  {\it Rev. Mod. Phys.} {\bf 81} (2009) 109.
\bibitem{Hasan}M. Z. Hasan and C. L. Kane, Rev. Mod. Phys. {\bf 82}, 3045 (2010).
\bibitem{Wilczek}F. Wilczek, {\it Fractional Statistics and Anyon Superconductivity}, (World Scientific, Singapore, 1990).
\bibitem{Iengo}R. Iengo and K. Lechner, {\it Anyon Quantum Mechanics and Chern-Simons theory}, Phys. Rep. {\bf 213}, 179 (1992).
\bibitem{Pol01}A. M. Polyakov, Mod. Phys. Lett. {\bf 2}, 893 (1987). 
\bibitem{W1}E. Witten, {\it Topological Quantum Field Theory}, Commun. Math. Phys. {\bf 117}, 353 (1988). 
\bibitem{W2}E. Witten, {\it Quantum Field Theory and the Jones Polynomial}, Commun. Math. Phys. {\bf 121}, 351 (1989).
\bibitem{W3}E. Witten, {\it Chern–Simons Theory as a String Theory}, Prog. Math. {\bf 133}, 637 (1995).
\bibitem{Marcos}M. Marino, {\it Chern–Simons Theory and Topological Strings}, Rev. Mod. Phys. {\bf 77}, 675 (2005).
\bibitem{Seiberg}S. Elitzur, G. Moore, A. Schwimmer and N. Seiberg, {\it Remarks on the Canonical Quantization of the Chern-Simons-Witten Theory}, Nucl. Phys. B {\bf 326}, 108 (1989).
\bibitem{Lab}J. Labastida and A. Ramallo, {\it Chern-Simons Theory and Conformal Blocks}, Phys. Lett. B {\bf 228}, 214 (1989).
\bibitem{Poly}A. Polychronakos, {\it Abelian Chern-Simons Theories in 2+1 Dimensions}, Ann. Phys. {\bf 203}, 231 (1990).
\bibitem{Nair}M. Bos and V. P. Nair, {\it Coherent State Quantization of Chern-Simon Theory}, Int. J. Mod. Phys. A {\bf 5}, 959 (1990).



\bibitem{H02}C. R. Hagen, Phys. Rev. D {\bf 31}, 848 (1985).
\bibitem{BDP}K. S. Babu, A. Das and P. K. Panigrahi, Phys. Rev. D {\bf 36}, 3725 (1987).
\bibitem{Nakamura}K. Nakamura {\it et. al.}, J. Phys. G {\bf 37}, 075021 (2010) 114.

\bibitem{Dunne}G. V. Dunne, {\em Lectures on Topological Aspects of Lower Dimensional Systems: 1998 Les Hausches Summer School}, arXiv:hep-th/9902115.
%\bibitem{JP}R. Jackiw and S.-Y. Pi, Phys. Rev. Lett. {\bf 98}, 266402 (2007).
\bibitem{Own}K. Abhinav, V. M. Vyas and P. K. Panigrahi, Pramana-journal of physics {\bf 85}, 1023 (2015).
\bibitem{V1}P. K. Panigrahi, V. M. Vyas and T. Shreecharan, arXiv:0901.1034[cond-mat.supr-con]; V. M. Vyas and P. K. Panigrahi, arXiv:1107.5521[cond-mat.supr-con].
\bibitem{P}A. M. Polyakov, Mod. Phys. Lett. {\bf 2}, 893 (1987); V. G. Knizhnik, A. M. Polyakov and A. B. Zamolodchikov, Mod. Phys. Lett. A {\bf 3}, 819 (1988); A. H. Chamseddine and M. Reuter, Nucl. Phys. B {\bf 317}, 757 (1989); A. M. Polyakov, Int. J. Mod. Phys. A {\bf 5}, 833 (1990).
\bibitem{D}A Das, W-J Huang and S Roy, Int. J. Mod. Phys. 7, 3447 (1992).
\bibitem{W}E. Witten, Comm. Math. Phys. {\bf 117}, 353 (1988); {\it ibid.}, Nucl. Phys. B {\it 311}, 46 (1988/1989).
\bibitem{Das}A. Das, {\it Field Theory: A Path Integral Approach}, (World Scientific, Singapore, 2006).
\bibitem{CFT1}A. A. Belavin, A. M. Polyakov and A. B. Zamolodchikov, Nucl. Phys. B {\bf 241}, 333 (1984). 
\bibitem{CFT2}P. Ginsparg, {\it Applied Conformal Field Theory}, Les Houches, Session XLIX, 1988, Champs, Fields, Strings and Critical Phenomena, ed. by E. Br\'ezin and J. Zinn-Justin, Elsevier Science Publishers B.V. (1989), arXiv:hep-th/9108028.
\bibitem{CCJ}C. Callan, S. Coleman and R. Jackiw, Ann. Phys. {\bf 59}, 42 (1970).
%\bibitem{Deser}S. Deser, Ann. Phys. {\bf 59}, 248 (1970).
\bibitem{Fujikawa}K. Fujikawa and H. Suzuki, {\em Path Integrals and Quantum Anomalies}, Clarendon Press, Oxford, (2004).
\end{thebibliography}
\end{document}